\documentclass[aps,prb,secnumarabic, bibnotes, twocolumn,superscriptaddress]{revtex4-1}
\usepackage{amsfonts}
\usepackage{mathrsfs}
\usepackage{amsmath}% needed for subequations
\usepackage{color}
\usepackage{natbib}
\usepackage{graphicx}
\usepackage{bm}% bold maths
\usepackage{amssymb}
\usepackage{xspace}
\usepackage{epstopdf}
\usepackage{dcolumn}% Align table columns on decimal point
\usepackage{longtable}
\usepackage{siunitx} %%for angles
\usepackage{multirow}
\usepackage[colorlinks=true, letterpaper=true, pdfstartview=FitV, linkcolor=blue, citecolor=blue, urlcolor=blue]{hyperref}

\makeatletter

\newcommand{\Rmnum}[1]{\expandafter\@slowromancap\romannumeral #1@}
\makeatother
\begin{document}

\title{Doping induced itinerant ferromagnetism and enhanced ferroelectricity in BL-InSe}

\author{Junlan Shi}
%\email[]{2133673001@qq.com}
\affiliation{College of Physics and Electronic Engineering, Center for Computational Sciences, Sichuan Normal University, Chengdu, 610068, China}

\author{Li Chen}
%\email[]{2133673001@qq.com}
\affiliation{College of Physics and Electronic Engineering, Center for Computational Sciences, Sichuan Normal University, Chengdu, 610068, China}

\author{Jiani Zhang}
%\email[]{2133673001@qq.com}
\affiliation{College of Physics and Electronic Engineering, Center for Computational Sciences, Sichuan Normal University, Chengdu, 610068, China}

\author{Botao Fu}
\email[]{fubotao2008@gmail.com}
\affiliation{College of Physics and Electronic Engineering, Center for Computational Sciences, Sichuan Normal University, Chengdu, 610068, China}

\date{\today}

\begin{abstract}
The microscopic coexistence of ferroelectricity and ferromagnetism in solids remains a fundamental challenge in condensed matter physics, with far-reaching implications for multifunctional materials and next-generation electronic devices. Using first-principles calculations, we predict emergent sliding ferroelectricity and doping-mediated ferromagnetism in bilayer (BL) InSe. The energetically favored AB stacked BL-InSe spontaneously breaks the out-of-plane mirror symmetry, resulting in a switchable polarization with a saturated component of 0.089 pC/m and a low transition barrier of 28.8 meV per unit cell. Strikingly, low-concentration electrostatic doping enhances rather than suppresses the ferroelectric polarization due to the abnormal layer-dependent electronic occupation in BL-InSe, in contrast to the conventional screening paradigm. In addition, the characteristic Mexican-hat-shaped valence band enables doping-induced itinerant half-metallic ferromagnetism, where the interlayer spin density difference scales linearly with doping concentration and can be reversed by switching the polarization direction. These results demonstrate the coexistence of ferroelectric and ferromagnetic orders in BL-InSe and establish a viable platform for realizing voltage-tunable multiferroicity through stacking and carrier doping in otherwise nonpolar and nonmagnetic semiconductors.

\end{abstract}

\maketitle

\section{Introduction}

The pursuit of multiferroic materials that simultaneously host ferroelectricity and ferromagnetism represents a fundamental frontier in condensed matter physics, driven by their potential to enable magnetoelectric coupling in next-generation devices\cite{chen2023magnetoelectric,eerenstein2006multiferroic,ortega2015multifunctional,mostovoy2024multiferroics,xu2020electric}. However, conventional magnetoelectric multiferroics remain scarce, primarily due to the intrinsic incompatibility between the mechanisms that govern the ferroelectric and ferromagnetic ordering\cite{spaldin2010multiferroics}. First, ferroelectric polarization typically requires empty orbitals $d/f$ to facilitate ionic displacement, while magnetism requires partially filled orbitals $d/f$ to establish spin alignment\cite{hill2000there}. Second, the metallic conductivity inherent to ferromagnets often screens electric polarization through free carrier redistribution. Since the discovery of the magnetoelectric effect in Cr$_2$O$_3$\cite{rado1962magnetoelectric,PhysRevB.86.094430}, these dual constraints have made intrinsic magnetoelectric multiferroics exceptionally rare.

In 2017, Wu \textit{et al.} proposed the concept of sliding ferroelectricity\cite{li2017binary}, which greatly expanded the landscape of two-dimensional (2D) ferroelectrics\cite{PhysRevLett.130.146801,wu2021sliding}. This mechanism reveals that artificially stacking nonpolar 2D monolayers can break the centrosymmetry of pristine layers and induce a sizable out-of-plane (OOP) ferroelectric polarization. This stacking-driven mechanism has since been demonstrated theoretically and experimentally in a range of 2D systems, including $h$-BN\cite{PhysRevB.106.054104,moore2021nanoscale,yasuda2021stacking,PhysRevMaterials.8.044001} and transition-metal dichalcogenides\cite{wang2022interfacial,weston2022interfacial,meng2022sliding}, and has been generalized to multilayers elemental materials\cite{wang2018two,PhysRevB.110.125434,liang2021out} and quasi-one-dimensional systems\cite{PhysRevB.110.024115}.
The resulting paradigm, engineering polarity through stacking rather than through $d/f$ orbital chemistry, substantially expands the design space for low-dimensional ferroelectrics~\cite{PhysRevB.111.L201406,liu2024Fatigue}.

The stacking route also decouples ferroelectricity from conventional insulating requirements. In weakly coupled van der Waals materials, anisotropic screening can allow an OOP polarization to coexist with in-plane metallicity~\cite{PhysRevB.91.064104}. Such ferroelectric metals have been reported experimentally (notably WTe$_2$) and explored theoretically in other systems with strong spin–orbit and topological characters~\cite{PhysRevLett.133.186801,fei2018ferroelectric,sharma2019room,yang2018origin,orlova2021evidence}. These developments suggest that stacking and carrier engineering together offer routes to integrate ferroelectric, metallic and topological orders in atomically thin platforms.

Achieving multiferroicity in two dimensions remains challenging. Most existing strategies mainly follow two routes: (i) stacking magnetic layers to break inversion symmetry and induce sliding ferroelectricity~\cite{PhysRevLett.133.246703,PhysRevB.110.224418,PhysRevLett.134.216801,xun2024coexisting,wu2024coexistence,hu2024ferrielectricity,li2024observation}, or (ii) starting from intrinsically ferroelectric systems and introducing magnetism by doping or heterostructure design~\cite{yang2020iron,park2025coexisting}. Stacking-engineered responses have been predicted in BL-CrI$_3$\cite{bwwv-f247} and BL-VS$_2$\cite{PhysRevLett.125.247601}, while hole doping in ferroelectric semiconductors such as monolayer In$_2$Se$_3$\cite{liu2021tunable} can drive itinerant ferromagnetism near van Hove singularities~\cite{song2022evidence,liu2019gamma}. However, both approaches rely on materials with pre-existing magnetism or polarity, limiting candidate space and design flexibility.

Based on this background, we propose a general strategy for designing multiferroic materials: nonmagnetic and nonpolar 2D monolayers with unique band structures are used as building blocks, where interlayer sliding breaks mirror symmetry ($M_z$) to induce OOP ferroelectric polarization, and hole doping introduces itinerant ferromagnetism, thereby enabling the coexistence of both orders in a single system. 
To validate this concept, we select experimentally stable InSe monolayers as a model system. 
First-principles calculations reveal that AB stacking BL-InSe spontaneously breaks the $M_z$ mirror, generating a switchable vertical polarization ($P_z$ = 0.089 pC/m) with a low transition barrier of 28.9 meV per unit cell. Counterintuitively, low-concentration hole doping enhances the polarization strength through interlayer charge asymmetry, in contrast to the conventional screening picture. Meanwhile, the characteristic ``Mexican-hat'' valence band leads to a van Hove singularity, giving rise to doping-induced itinerant ferromagnetism with magnetic moments saturating at 1.0 $\mu_B$ per hole and a linear scaling of interlayer spin asymmetry with doping concentration. We further show that carrier doping acts as a dual control knob to simultaneously tune ferroelectric and magnetic order parameters, offering a promising route toward voltage-tunable multiferroic phases in engineered quantum materials.

\section{Computational methods}
All first-principles calculations were performed within density functional theory (DFT) as implemented in the Vienna Ab initio Simulation Package (VASP) \cite{PhysRev.136.B864,PhysRev.140.A1133,KRESSE199615}. Exchange–correlation effects were treated using the generalized-gradient approximation (GGA) with the Perdew–Burke–Ernzerhof (PBE) functional \cite{PhysRevB.54.11169,PhysRevLett.77.3865}. In addition, the screened hybrid Heyd–Scuseria–Ernzerhof functional (HSE06)\cite{paier2006screened} was employed to obtain more accurate electronic structures. After convergence tests, a plane-wave cutoff energy of 400 eV and a vacuum thickness of 26~{\AA} were used.  
To properly account for the interlayer van der Waals (vdW) interactions in BL-InSe, we systematically tested different vdW correction schemes [Fig.~S1 and Tab.~S1 in Supplementary Materials (SM)] and identified the optB86-vdW functional~\cite{PhysRevB.83.195131} as providing the best agreement with experimental results.
Brillouin-zone integrations were carried out using a $\Gamma$-centered $21\times21\times1$ Monkhorst–Pack $k$-point mesh \cite{PhysRevB.40.3616}. Total energies were converged to within $10^{-5}$ eV and all atomic forces were relaxed below $10^{-2}$ eV \AA$^{-1}$. Phonon dispersion relations were computed using the Phonopy package\cite{ALFE20092622} by means of density-functional perturbation theory (DFPT) on a $4{\times}4{\times}1$ supercell.

%%%%%%%%%%%%%%
\section{ RESULTS AND DISCUSSION}
\subsection{Stacking Landscape of BL-InSe} 
\begin{figure}
	\includegraphics[width=8.5cm]{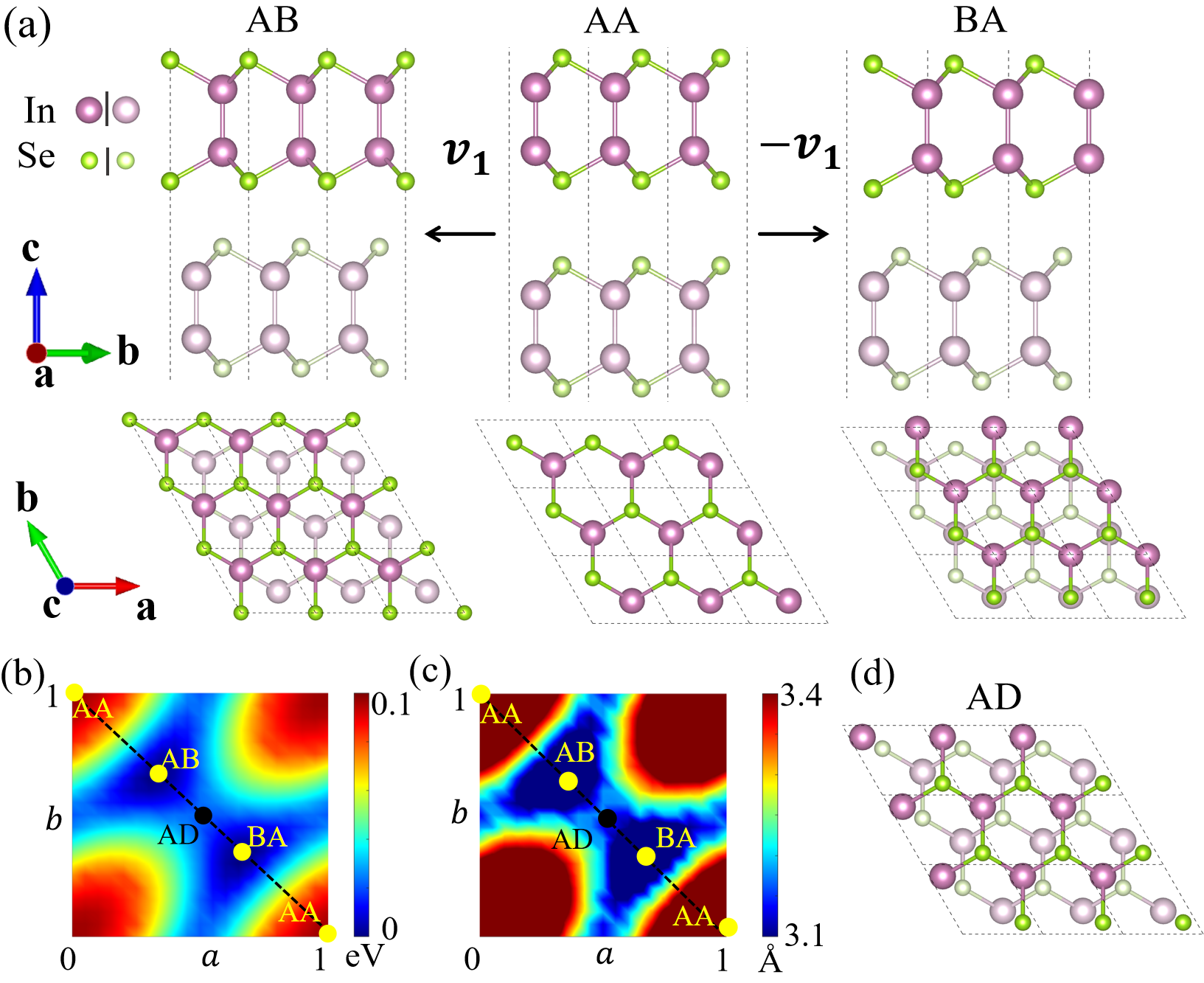}
    \caption{ (a) Side and top views of BL-InSe in AB, AA, and BA stackings with arrows indicating the interlayer sliding vectors. (b)-(c) Interlayer binding energy and interlayer distance versus sliding displacement. (d) Top view of AD stacked BL-InSe.}\label{Fig1}
\end{figure}

\begin{figure*}
	\includegraphics[width=17.6cm]{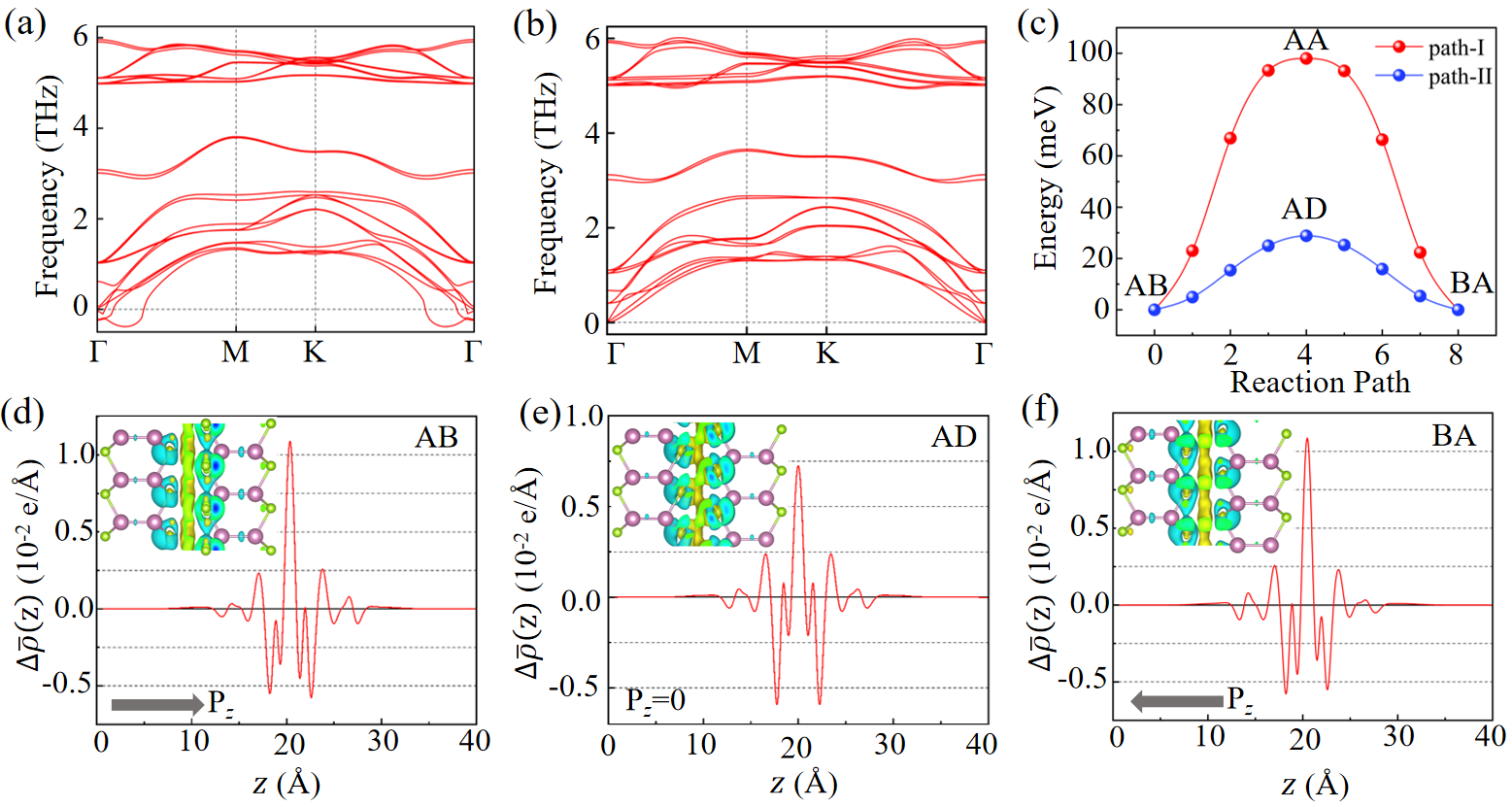}
    \caption{ (a)-(b) Phonon spectra of the AA and AB stackings.
(c) Energy profile of ferroelectric switching as a function of the number of CI-NEB steps. (d)-(f) Differential charge density and planar-averaged charge density maps for the AB, AD, and BA phases, respectively. The yellow and cyan regions denote charge accumulation and depletion.}\label{Fig2}
\end{figure*}

Layered bulk InSe, a van der Waals semiconductor with ultrahigh carrier mobility and favorable optoelectronic properties, has emerged as a versatile platform for multifunctional devices due to its phase-tunable polarization\cite{bandurin2017high,zhao2023probing,li2018high,sun2018inse,li2024enhancement}. It is noteworthy that its distinctive stacking configurations ($\beta$-, $\varepsilon$-, and $\gamma$-phases) govern diverse ferroelectric behaviors: the $\beta$-phase exhibits robust bidirectional polarization switching at room temperature\cite{hu2019room}, the $\varepsilon$-phase shows layer-parity-modulated polarization\cite{wang2024sliding}, and the $\gamma$-phase enables extrinsic tunability via rare-earth doping\cite{sui2023sliding}. Motivated by these phase-dependent ferroic behaviors in bulk, we next consider the monolayer limit, where dimensional reduction suppresses interlayer interactions while preserving lattice symmetry and electronic features.
Monolayer (ML) InSe has been successfully synthesized by chemical vapor deposition (CVD)~\cite{chang2018synthesis}. Its structure consists of vertically stacked Se–In–In–Se quadruple layers, characterized by a threefold rotational axis ($C_3$) and a horizontal mirror plane ($M_z$), which together suppress OOP ferroelectricity.
Previous theoretical studies have shown that ML-InSe exhibits broad optical absorption and high mechanical flexibility\cite{gao2021monolayer,luo2020ab}. In particular, its Mexican-hat-shaped valence band provides a fertile platform for emergent phenomena under hole doping, including ferromagnetism, high-temperature superconductivity, and charge density waves (CDWs)~\cite{fu2017effects,PhysRevB.99.085409,huang2024complex}.

%%%
Building on the properties of ML-InSe, we next investigate BL-InSe, where interlayer stacking provides an additional degree of freedom for engineering ferroic orders. 
Within the hexagonal lattice, six high-symmetry bilayer configurations~\cite{yang2017electric} are possible and can be grouped into centrosymmetric parallel and non-centrosymmetric 60$^\circ$-rotated stackings~\cite{stern2010interfacial}. Here we focus on the parallel stackings—AA, AB, and BA—because of their polar nature and potential for ferroic control [Fig.~\ref{Fig1}(a)].
Among them, AB and BA are inversion-related structures, both belonging to space group $P3m1$ (No.~156) with polar point group $C_{3v}$, forming an energetically degenerate bistable ferroic pair. In contrast, AA adopts space group $P\overline{6}m2$ (No.~187) with a nonpolar point group $D_{3h}$. Importantly, these three configurations are connected by interlayer sliding: translating the top layer by $\pm \textbf{\textit{v}}_1 = \pm \tfrac{1}{3}(-\textbf{\textit{a}}+\textbf{\textit{b}})$ transforms AA into AB or BA, respectively.

%%%
To evaluate structural stability, we computed the stacking energy landscape by sliding the top layer from the $M_z$-preserved AA phase [Fig.~\ref{Fig1}(b)]. The profile exhibits two minima (AB and BA phases) and a maximum at AA phase. Phonon spectra support this result: AA shows imaginary modes near the $\Gamma$ point [Fig.~\ref{Fig2}(a)] indicating is dynamical instability, whereas AB is dynamically stable [Fig.~\ref{Fig2}(b)]. A saddle point, denoted AD [Fig.~\ref{Fig1}(d)], lies between AB and BA, belonging to the non-symmorphic space group $Abm2$ (No.~39) and connecting the two phases via the slip vector $\textbf{\textit{v}}_2=\tfrac{1}{2}\textbf{\textit{v}}_1$. Interlayer distances [Fig.~\ref{Fig1}(c)] correlate directly with energy: AA has the largest spacing, while AB/BA have the smallest, consistent with their stability.

\subsection{Sliding Ferroelectricity in BL-InSe}
\begin{figure*}
	\includegraphics[width=17.6cm]{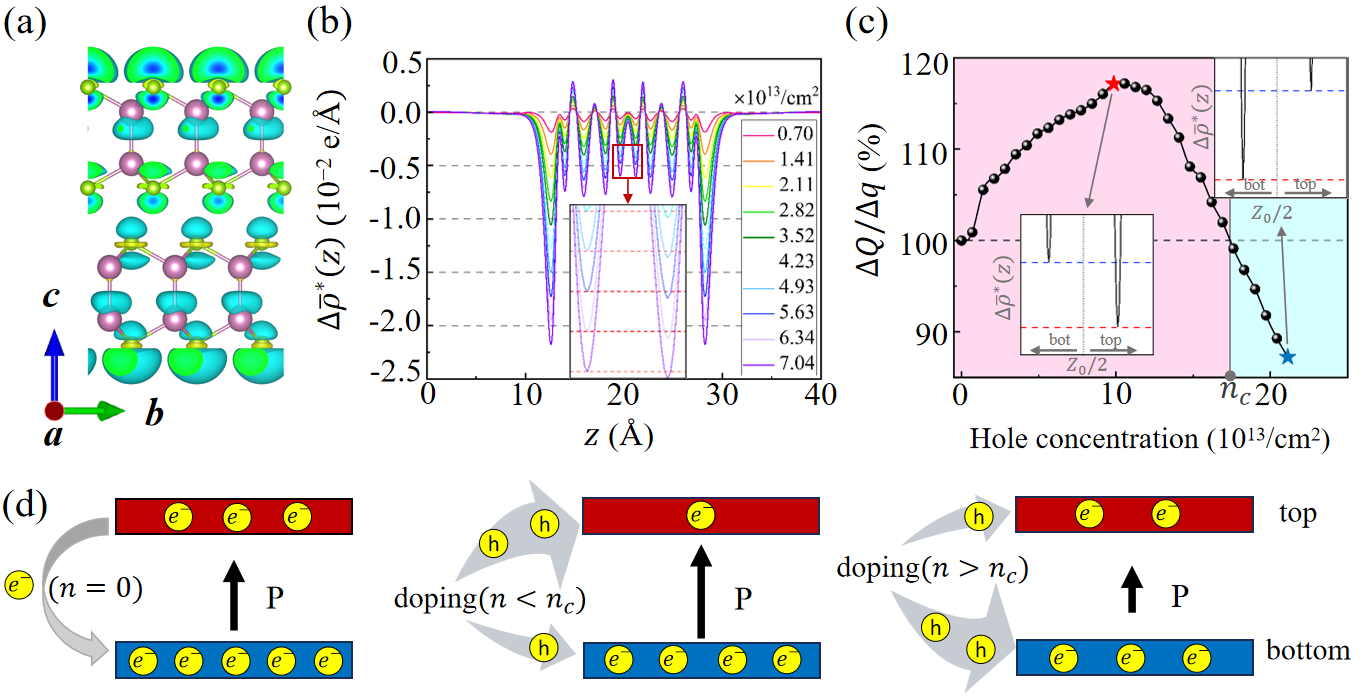}
\caption{ (a) Differential charge density between doped and undoped BL-InSe at $n$=7.04$\times$10$^{13}$ cm$^{-2}$. (b) The planar-averaged charge density along $z$ for various doping levels. (c) The relative polarization versus hole concentration with an inset highlighting the differential charge near $Z_0/2$ at the two marked points ($n_c = 17.39 \times 10^{13},\text{cm}^{-2}$). 
(d) Schematic diagram illustrating the effect of hole doping on OOP polarization.}\label{Fig3}
\end{figure*}

%%%%
In the AB and BA stacking configurations, the broken $M_z$ symmetry induces spontaneous OOP polarization. Using the Berry phase method~\cite{PhysRevB.47.1651,RevModPhys.66.899}, we obtained a ferroelectric polarization of $P_z = \mp 0.089$~pC/m for AB and BA, which is slightly below the literature value because of the distinct vdW correction~\cite{li2017binary}. To investigate the polarization switching mechanism, we performed climbing-image nudged elastic band (CI-NEB) simulations~\cite{henkelman2000climbing}, which reveal two distinct pathways [Fig.~\ref{Fig2}(c)]: path-I through the paraelectric AA phase with a high barrier of 98.02~meV, and path-II via the intermediate AD phase with a much lower barrier of 28.86~meV. 
The latter thus represents the energetically favorable switching route. Compared with previous studies on sliding ferroelectrics in 2D systems~\cite{PhysRevB.110.205119}, this low barrier underscores the intrinsic efficiency of AD-mediated polarization reversal in BL-InSe, highlighting its potential for low-power ferroelectric devices.

To further elucidate the microscopic origin of OOP ferroelectricity in BL-InSe, we computed the differential charge density and its planar average integrated along the $z$-axis to analyze interlayer charge redistribution\cite{liu2019vertical}. The differential charge density is defined as
\begin{equation}
    \Delta \rho(r) = \rho_{\text{BL}}(r) - \rho_{\text{top}}(r) - \rho_{\text{bot}}(r),
\end{equation}
where $\rho_{\text{BL}}(r)$, $\rho_{\text{top}}(r)$, and $\rho_{\text{bot}}(r)$ are the charge densities of the bilayer system, the isolated top, and isolated bottom monolayers, respectively~\cite{WANG2021108033}. The planar average differential charge density is given by
\begin{equation}
    \Delta \overline{\rho}(z) = \int\Delta\rho(r)\,dxdy,
\end{equation}
As shown in Fig.~\ref{Fig2}(d)-(f), all three stackings (AD, AB, and BA) exhibit electron depletion near the middle Se layer and accumulation in the interlayer region. Compared with the AD stacking, the AB and BA display an asymmetric distribution between the top and bottom layers, more clearly seen in the planar averages.
To quantify this effect, we calculated the interlayer charge transfer as
\begin{equation}
    \Delta q = q_{\text{bot}} - q_{\text{top}},
\end{equation}
where $q_{\text{bot}} = \int_{0}^{Z_0/2} \overline{\rho}(z)dz$ and $q_{\text{top}} = \int_{Z_0/2}^{Z_0} \overline{\rho}(z)dz$. The calculated $\Delta q$ = 0.0071~$e$ indicates charge transfer from the top to the bottom layer in the AB stacking, generating a local dipole moment that leads to upward OOP polarization as illustrated on the left panel of Fig.~\ref{Fig3}(d). Conversely, BA shows the opposite charge transfer, leading to downward polarization. By contrast, the AD stacking shows symmetric charge distribution, consistent with its vanishing polarization.

\subsection{Tunable Ferroelectricity via Electrostatic Doping}

Ferroelectricity in 2D semiconductors is governed by the subtle balance between intrinsic lattice asymmetry and electrically induced carrier redistribution\cite{cui2018two}.
To investigate how hole doping (see the electron doping in Fig.~S2 in SM) modulates the OOP polarization of AB-stacking BL-InSe, we first analyze the charge rearrangement upon hole doping. 
The differential charge density between the doped ($\rho_{\text{doped}}$) and undoped ($\rho_{\text{undoped}}$) systems is defined as
\begin{equation}
\Delta\rho^{*}(r)=\rho_{\text{doped}}(r)-\rho_{\text{undoped}}(r).
\end{equation}
Owing to the intrinsic polar discontinuity, $\Delta\rho^{*}(r)$ is asymmetric and develops a weak interlayer imbalance with increasing carrier density [Fig.~\ref{Fig3}(a)]. To better visualize this imbalance, we computed the planar-averaged differential charge density $\Delta\overline{\rho}^{*}(z)$, analogous to Eq.~(2). As shown in Fig.~\ref{Fig3}(b), $\Delta\overline{\rho}^{*}(z)$ highlights the unequal distribution of injected carriers between the two layers. This asymmetric distribution is expected to directly feed back into the macroscopic polarization.

Because the doped system deviates from charge neutrality, the conventional Berry-phase method is no longer applicable. Instead, we characterize the polarization indirectly using the interlayer charge transfer parameter $\Delta Q$, defined as
\begin{equation}
\Delta Q=\Delta q+\Delta q^*,
\end{equation}
where $\Delta q$ is the intrinsic charge transfer caused by the non-centrosymmetric stacking (responsible for $P_{z}$), and $\Delta q^*$ is the additional transfer induced by the doped holes (polarization increment $P^*_z$). Within the linear-response regime, the relative polarization is then simply
\begin{equation}
\frac{P_z+P^*_z}{P_{z}} \approx \frac{\Delta Q}{\Delta q}=1+\frac{\Delta q^*}{\Delta q}.
\end{equation}

The calculated polarization ratio as a function of hole density $n$ is presented in Fig.~\ref{Fig3}(c).
At low doping ($n< n_c$), the ratio exceeds $100\%$ and peaks at $n=9.85\times10^{13}$~cm$^{-2}$, indicating an anomalous enhancement of ferroelectricity that defies the conventional screening paradigm~\cite{PhysRevB.109.235426}. 
To clarify this enhancement, we examine the planar-averaged differential charge density at two representative doping levels (insets in Fig.~\ref{Fig3}(c)). Specifically, at low doping ($n<n_c$), the doped holes preferentially occupy the top layer, thereby enhancing the interlayer asymmetry and amplifying the polarization. In contrast, once the doping exceeds the critical density ($n>n_c$), the holes shift to the bottom layer, which reduces the asymmetry and consequently suppresses the polarization below the intrinsic value.

Microscopically, this anomalous behavior originates from the abnormal layer-dependent electronic occupation in the band structure of BL-InSe. As confirmed by the layer-resolved band structure in Fig.~S3 in SM, the valence-band maximum is predominantly localized on the top layer.
Consequently, as depicted in Fig.~\ref{Fig3}(d), the initially injected holes preferentially occupy the top layer more than bottom layer ($\Delta q^* \Delta q > 0$), thereby reinforcing the intrinsic dipole. However, once the top-layer states become saturated, additional holes start to populate the lower layer. This occupation reverses the sign of $\Delta q^*$, leading to a reduction of the interlayer asymmetry and ultimately suppressing the polarization below its intrinsic value.

We note that carrier-enhanced ferroelectricity has also been reported in ML-SnS and BL-PtTe$_2$-family systems, where the enhancement was attributed to abnormal occupation of bonding/antibonding states near the Fermi level~\cite{PhysRevB.108.104109,PhysRevB.111.094516}.
Compared with monolayers, bilayers provide an additional interlayer degree of freedom. The crossover from cooperative amplification to competitive suppression thus offers a distinct mechanism and a quantitative roadmap for engineering and reversibly tuning OOP ferroelectricity in 2D semiconductors.

\subsection{Doping-Induced Itinerant Ferromagnetism}
\begin{figure}
 \includegraphics[width=8.6cm]{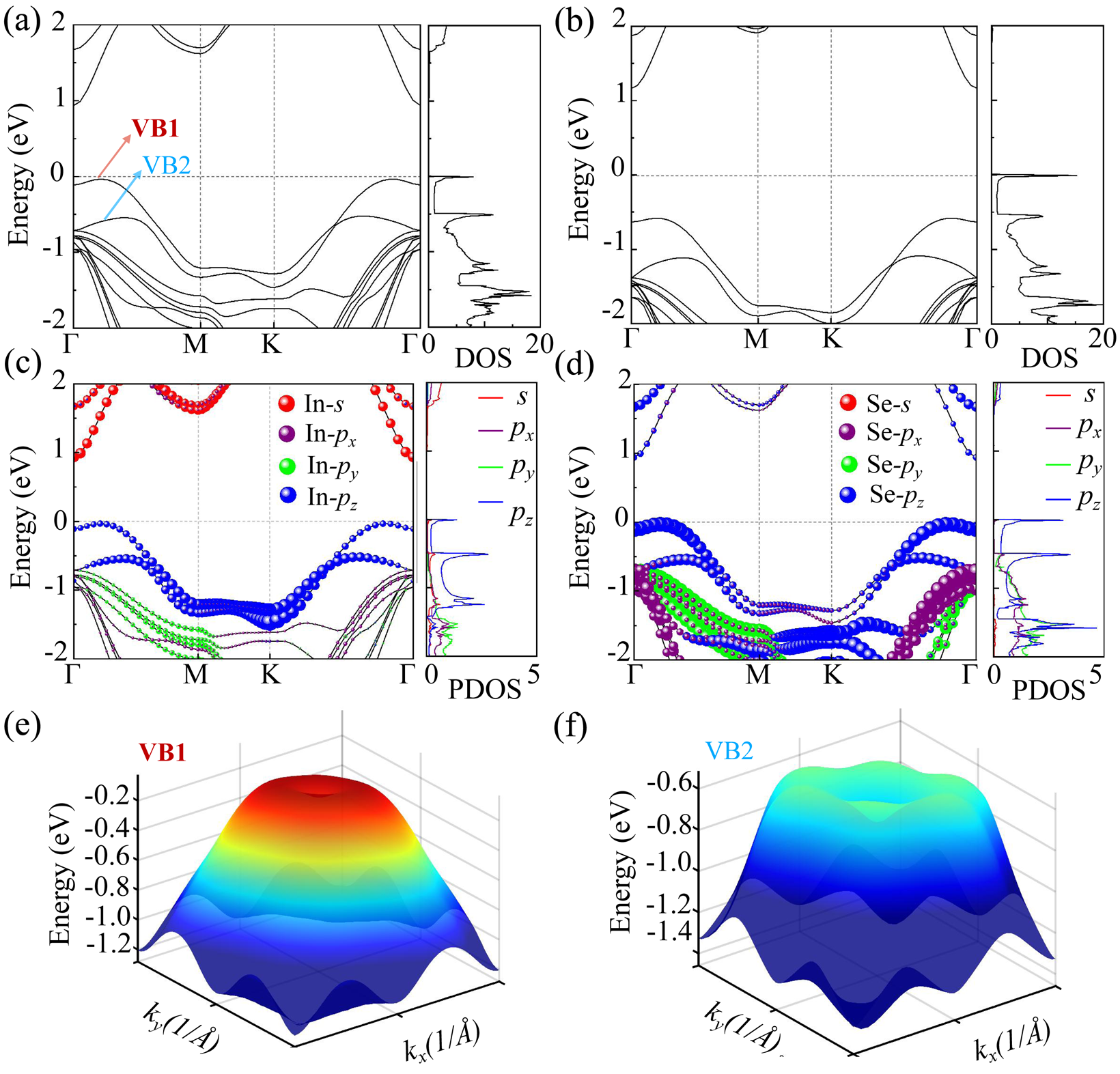}
  \caption{(a) and (b) show the band structure and DOS of the AB-stacking BL-InSe using the PBE functional and the HSE06 method, respectively. (c) and (d) present the band projections and PDOS for In and Se atoms in the of BL-InSe, respectively. (e) and (f) depict the three-dimensional Mexican hat-shaped band structures corresponding to VB1 amd VB2 in (a).}\label{Fig4}
\end{figure}

\begin{figure}[bth]
\includegraphics[width=8.6cm]{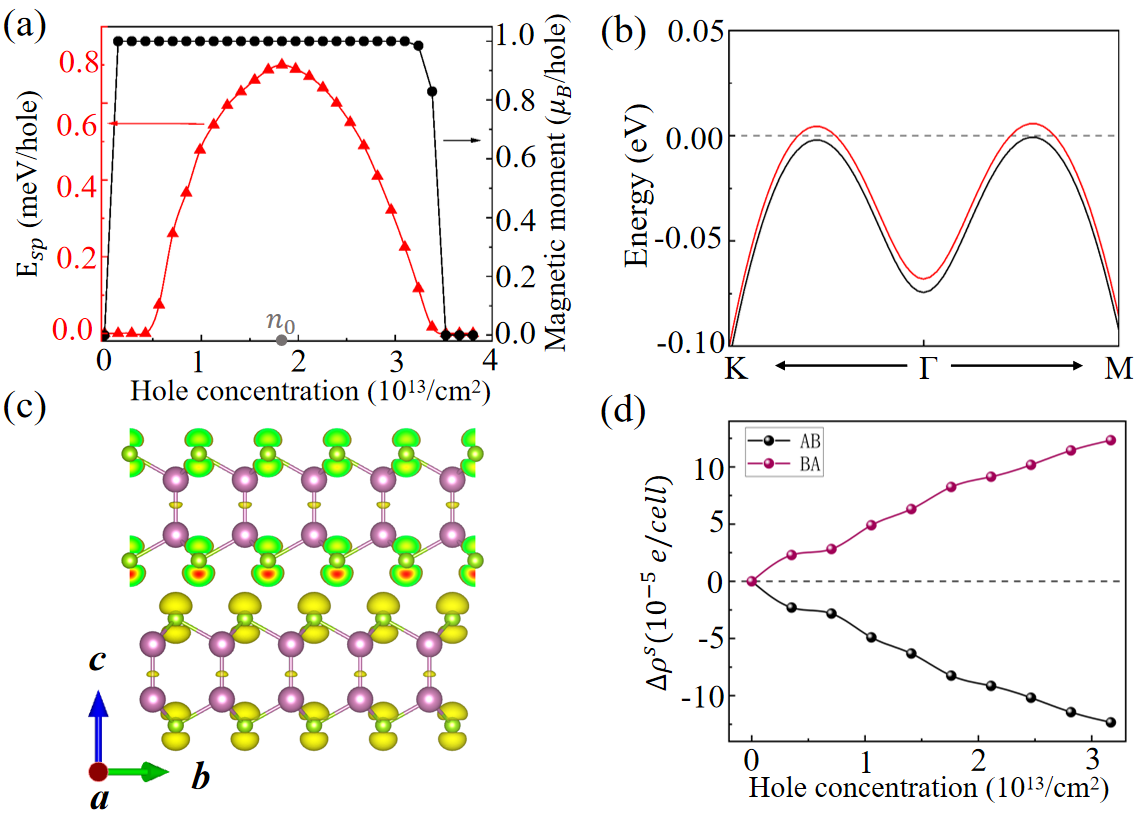}
  \caption{(a) Spin magnetic moment and spin-polarization energy of AB-stacked BL-InSe as functions of hole doping concentration. (b) Spin-polarized band structure at $n_0=1.83\times 10^{13}\text{cm}^{-2}$, with the corresponding spin charge density distribution shown in (c). (d) Interlayer spin charge difference as a function of hole doping concentration for AB and BA stackings}\label{Fig5}
\end{figure}

To benchmark our study, we first revisited ML-InSe, whose electronic structure and doping-induced magnetism have been reported in previous works\cite{PhysRevLett.114.236602,houssa2021doping}. Our first-principles calculations (see FIG.~S4 in SM) confirm that ML-InSe is an indirect bandgap semiconductor with a characteristic Mexican-hat dispersion near the $\Gamma$ point. The valence band maximum (VBM) is mainly derived from the $p_z$ orbitals of Se, while the conduction band minimum (CBM) originates from $s$ orbital of In. This band structure produces a sharp DOS peak close to the VBM, where carrier doping may trigger itinerant ferromagnetism\cite{PhysRevB.102.195408,lin2017magnetism,cai2025non,li2023spin}. Consistently, our calculations reproduce the well-established result that small hole doping induces a spin moment, which quickly saturates at $1.0\,\mu_B$/hole. These findings provide a reliable basis for extending doping studies to bilayer systems.

%%%%%%%%%%%%
%%%%%%%%%%%%%%%%%%%%
We next focus on BL-InSe, where interlayer stacking modifies the electronic structure and offers new opportunities to engineer ferroic and magnetic orders. For the AB stacking configuration, band structure and DOS calculations using both PBE and HSE06 functionals [Figs.~\ref{Fig4}(a)–(b)] reveal similar dispersions but different indirect bandgaps (0.97 eV vs. 1.75 eV). Compared with the monolayer, AB stacking bilayer splits the VBM into two subbands (VB1 and VB2), both with Mexican-hat dispersions [Figs.~\ref{Fig4}(e)–(f)]. This splitting transforms the monolayer’s single DOS peak into a characteristic double-peak structure. Projected band analysis further shows that both VB1 and VB2 mainly arise from In and Se $p_z$ orbitals, with Se $p_z$ states dominating around $\Gamma$ [Figs.~\ref{Fig4}(c)–(d)].

Introducing hole doping near the first DOS peak closest to the Fermi level, we uncover robust itinerant ferromagnetism in BL-InSe due to Stoner criterion~\cite{PhysRevLett.114.236602}. As shown in Fig.~\ref{Fig5}(a), the magnetic moment rapidly increases and saturates at $1.0~\mu_B$/hole within a broad doping window ($0.14$–$3.10\times10^{13}$ cm$^{-2}$) before disappearing at higher concentrations. The spin-polarization energy $E_{sp}$, defined as the energy difference between the ferromagnetic (FM) and nonmagnetic (NM) states, remains positive across the entire range of doping levels, indicating a pronounced energetic preference for the ferromagnetic state, consistent with the monolayer case~\cite{houssa2021doping}. Notably, the saturation magnetic
moment of 1.0 $\mu_B$ per hole indicates a $100\%$ spin polarization, confirming the presence of half-metallic states [Fig.~\ref{Fig5}(b)]].

It is worthwhile to note that, within this doping range, hole injection not only induces robust ferromagnetism but also enhances the intrinsic sliding ferroelectricity. For instance, at an optimal concentration of $n_0$=1.83$\times$10$^{13}$ cm$^{-2}$, the electric polarization increases to 107$\%$, confirming the coexistence of ferroelectricity and ferromagnetism in BL-InSe. Microscopically, doping modifies the system polarization through asymmetric interlayer charge redistribution, which simultaneously affects the interlayer distribution of magnetism.
As shown in Fig.~\ref{Fig5}(c), the spin charge density distribution at $n_0$ reveals that the magnetic moments are primarily localized on the interfacial Se atoms and the outermost Se atoms, closely resembling the differential charge density map in Fig.~~\ref{Fig3}(a).
This effect is quantified by the interlayer spin-charge difference $\Delta q^s = q^s_{\text{bot}} -q^s_{\text{top}}$, which exhibits a nearly linear increase with hole concentration [Fig.~\ref{Fig5}(d)]. Reversing the stacking from AB to BA via an external electric field simultaneously flips the interlayer spin distribution, thereby enabling direct magnetoelectric coupling. These results demonstrate that BL-InSe provides a unique platform where electric-field-controlled ferroelectricity can reversibly manipulate spin polarization, offering a distinct pathway for low-power spintronic devices and non-volatile multi-state memory applications.

\section{summary}
%%%%%%%%%%%%%%%%
In summary, by taking experimentally synthesized ML-InSe as a prototype, we carried out first-principles calculations to explore the emergence of both ferroelectricity and ferromagnetism in its bilayer form. We find that the interlayer sliding in AB-stacked BL-InSe breaks the $M_z$ symmetry, giving rise to a spontaneous OOP polarization. Unlike conventional expectations where carrier doping typically screens polarization, our results reveal that low levels of both electron and hole doping unexpectedly enhance the ferroelectric polarization. This unusual behavior originates from the selective redistribution of carriers between layers and highlights a new route for carrier-controlled ferroelectricity. In addition, hole doping in BL-InSe, facilitated by its unique Mexican-hat valence band dispersion, induces ferromagnetic ordering through the sharp van Hove singularities at the valence band maximum. This dual emergence of doping-enhanced ferroelectricity and carrier-induced ferromagnetism illustrates a previously unexplored pathway to realize multiferroicity in two-dimensional systems. More broadly, our findings establish BL-InSe as a representative platform, and suggest that similar mechanisms may apply to a wide class of nonpolar and nonmagnetic 2D materials with comparable electronic structures, thereby opening new opportunities for designing magnetoelectric coupling in low-dimensional materials.

%%%%%%%%%%%%%%%%%%%%%%%%%%%%%%
%%%%%%%%%%%%%%%%%%%%%%%%%%%%%%%%
\begin{acknowledgements}
This research was funded by the National Natural Science Foundation of China (Grant No. 12204330) and the Key Project of Sichuan Science. We also thank the Sichuan Normal University for financial support (No. 341829001). The numerical computations were performed at the Hefei Advanced Computing Center, and this research was also supported by the High-Performance Computing Center of Sichuan Normal University, China.

\end{acknowledgements}

\bibliographystyle{apsrev4-2}

\end{document}